\documentclass[a4paper]{cernart}
\usepackage{graphicx}
\graphicspath{{figures/}}
\usepackage{rotating}

\usepackage{epsfig}
\usepackage{amsmath}
\usepackage{amssymb}
\usepackage[dvips]{color}
\usepackage{bm}
\newcommand{\la}{$\rm \Lambda$}
\newcommand{\al}{$\rm \bar \Lambda$}
\newcommand{\ks}{$\rm K^0_S$}

\begin {document}
\dimen\footins=\textheight

\begin{titlepage}
\docnum{CERN--PH--EP/2009--011}
\date{4 June 2009}
\vspace{1cm}

\title{\LARGE
Measurement of the Longitudinal Spin Transfer to\\
$\mathbf \Lambda$ and
$\mathbf {\bar{\Lambda}}$ Hyperons in Polarised Muon DIS
}
\vspace*{0.5cm}

\collaboration{COMPASS Collaboration}


\vspace{2cm}
\begin{abstract}
The longitudinal polarisation transfer from
muons to
\la~and \al~hyperons, $D_{LL}^{\Lambda(\bar{\Lambda})}$,  has been studied
in deep inelastic scattering off an unpolarised isoscalar target at the COMPASS experiment at CERN. The spin transfers to \la~and \al~produced in the current fragmentation region exhibit different behaviours as a function of $x$ and $x_F$. The measured $x$ and $x_F$ dependences of
$D_{LL}^{\Lambda}$ are compatible with zero, while  $D_{LL}^{\bar{\Lambda}}$ tends to increase with
$x_F$, reaching values of 0.4 - 0.5. The resulting average values are $D_{LL}^{\Lambda}$  = $-0.012 \pm 0.047 \pm 0.024$ and $D_{LL}^{\bar{\Lambda}}$ = $0.249\pm 0.056 \pm 0.049$. These results are discussed in the frame of recent model calculations.
\end{abstract}

\vspace*{60pt}


\vfill
\submitted{submitted to the European Physical Journal C}

\noindent
{{\large  COMPASS Collaboration}\\[\baselineskip]}
%
%
M.~Alekseev\Iref{turin_p},
V.Yu.~Alexakhin\Iref{dubna},
Yu.~Alexandrov\Iref{moscowlpi},
G.D.~Alexeev\Iref{dubna},
A.~Amoroso\Iref{turin_u},
A.~Austregesilo\IIref{cern}{munichtu},
B.~Bade{\l}ek\Iref{warsaw},
F.~Balestra\Iref{turin_u},
J.~Ball\Iref{saclay},
J.~Barth\Iref{bonnpi}
% ...
%
%

~
~
%
%
\Instfoot{bielefeld}{Universit\"at Bielefeld, Fakult\"at f\"ur Physik, 33501 Bielefeld, Germany\Aref{f}}
\Instfoot{bochum}{Universit\"at Bochum, Institut f\"ur Experimentalphysik, 44780 Bochum, Germany\Aref{f}}
\Instfoot{bonniskp}{Universit\"at Bonn, Helmholtz-Institut f\"ur  Strahlen- und Kernphysik, 53115 Bonn, Germany\Aref{f}}
\Instfoot{bonnpi}{Universit\"at Bonn, Physikalisches Institut, 53115 Bonn, Germany\Aref{f}}
\Instfoot{brno}{Institute of Scientific Instruments, AS CR, 61264 Brno, Czech Republic\Aref{g}}
\Instfoot{burdwan}{Burdwan University, Burdwan 713104, India\Aref{h}}
\Instfoot{calcutta}{Matrivani Institute of Experimental Research \& Education, Calcutta-700 030, India\Aref{i}}
\Instfoot{dubna}{Joint Institute for Nuclear Research, 141980 Dubna, Moscow region, Russia}
\Instfoot{erlangen}{Universit\"at Erlangen--N\"urnberg, Physikalisches Institut, 91054 Erlangen, Germany\Aref{f}}
\Instfoot{freiburg}{Universit\"at Freiburg, Physikalisches Institut, 79104 Freiburg, Germany\Aref{f}}
\Instfoot{cern}{CERN, 1211 Geneva 23, Switzerland}
\Instfoot{liberec}{Technical University in Liberec, 46117 Liberec, Czech Republic\Aref{g}}
\Instfoot{lisbon}{LIP, 1000-149 Lisbon, Portugal\Aref{j}}
\Instfoot{mainz}{Universit\"at Mainz, Institut f\"ur Kernphysik, 55099 Mainz, Germany\Aref{f}}
\Instfoot{miyazaki}{University of Miyazaki, Miyazaki 889-2192, Japan\Aref{k}}
\Instfoot{moscowlpi}{Lebedev Physical Institute, 119991 Moscow, Russia}
\Instfoot{munichlmu}{Ludwig-Maximilians-Universit\"at M\"unchen, Department f\"ur Physik, 80799 Munich, Germany\AAref{f}{l}}
\Instfoot{munichtu}{Technische Universit\"at M\"unchen, Physik Department, 85748 Garching, Germany\AAref{f}{l}}
\Instfoot{nagoya}{Nagoya University, 464 Nagoya, Japan\Aref{k}}
\Instfoot{praguecu}{Charles University, Faculty of Mathematics and Physics, 18000 Prague, Czech Republic\Aref{g}}
\Instfoot{praguectu}{Czech Technical University in Prague, 16636 Prague, Czech Republic\Aref{g}}
\Instfoot{protvino}{State Research Center of the Russian Federation, Institute for High Energy Physics, 142281 Protvino, Russia}
\Instfoot{saclay}{CEA DAPNIA/SPhN Saclay, 91191 Gif-sur-Yvette, France}
\Instfoot{telaviv}{Tel Aviv University, School of Physics and Astronomy, 69978 Tel Aviv, Israel\Aref{m}}
\Instfoot{triest_i}{Trieste Section of INFN, 34127 Trieste, Italy}
\Instfoot{triest}{University of Trieste, Department of Physics and Trieste Section of INFN, 34127 Trieste, Italy}
\Instfoot{triestictp}{Abdus Salam ICTP and Trieste Section of INFN, 34127 Trieste, Italy}
\Instfoot{turin_u}{University of Turin, Department of Physics and Torino Section of INFN, 10125 Turin, Italy}
\Instfoot{turin_i}{Torino Section of INFN, 10125 Turin, Italy}
\Instfoot{turin_p}{University of Eastern Piedmont, 1500 Alessandria,  and Torino Section of INFN, 10125 Turin, Italy}
\Instfoot{warsaw}{So{\l}tan Institute for Nuclear Studies and University of Warsaw, 00-681 Warsaw, Poland\Aref{n} }
\Instfoot{warsawtu}{Warsaw University of Technology, Institute of Radioelectronics, 00-665 Warsaw, Poland\Aref{o} }
\Instfoot{yamagata}{Yamagata University, Yamagata, 992-8510 Japan\Aref{k} }
%
%
\Anotfoot{+}{Deceased}
\Anotfoot{a}{Also at IST, Universidade T\'ecnica de Lisboa, Lisbon, Portugal}
\Anotfoot{b}{On leave of absence from JINR Dubna}
\Anotfoot{c}{Also at Chubu University, Kasugai, Aichi, 487-8501 Japan$^{\rm j)}$}
\Anotfoot{d}{Also at KEK, 1-1 Oho, Tsukuba, Ibaraki, 305-0801 Japan}
\Anotfoot{e}{Also at GSI mbH, Planckstr.\ 1, D-64291 Darmstadt, Germany}
\Anotfoot{f}{Supported by the German Bundesministerium f\"ur Bildung und Forschung}
\Anotfoot{g}{Suppported by Czech Republic MEYS grants ME492 and LA242}
\Anotfoot{h}{Supported by DST-FIST II grants, Govt. of India}
\Anotfoot{i}{Supported by  the Shailabala Biswas Education Trust}
\Anotfoot{j}{Supported by the Portuguese FCT - Funda\c{c}\~ao para a Ci\^encia e Tecnologia grants POCTI/FNU/49501/2002 and POCTI/FNU/50192/2003}
\Anotfoot{k}{Supported by the MEXT and the JSPS under the Grants No.18002006, No.20540299 and No.18540281; Daiko Foundation and Yamada Foundation}
\Anotfoot{l}{Supported by the DFG cluster of excellence `Origin and Structure of the Universe' (www.universe-cluster.de)}
\Anotfoot{m}{Supported by the Israel Science Foundation, founded by the Israel Academy of Sciences and Humanities}
\Anotfoot{n}{Supported by Ministry of Science and Higher Education grant 41/N-CERN/2007/0}
\Anotfoot{o}{Supported by KBN grant nr 134/E-365/SPUB-M/CERN/P-03/DZ299/2000}
\end{titlepage}
\section{\label{sec:introduction} Introduction}

 The study of the \la~and \al~hyperon polarisation in
DIS is important for the understanding of
the nucleon structure, the mechanisms of  hyperon production and the hyperon spin
structure.  In particular, it may
provide valuable information on the unpolarised strange quark
distributions $s(x)$ and $\bar{s}(x)$ in the nucleon. In this paper measurements of the longitudinal
polarisation of
\la~and \al~hyperons produced in deep-inelastic scattering
(DIS) of polarised muons off an unpolarised isoscalar target are
presented.

In a simple parton model a polarised lepton interacts
preferentially with a quark carrying a specific spin orientation.
After the absorption of the hard virtual photon
the remaining
system consists
of the scattered quark and the target remnant which are both
polarised. During the hadronisation, part of the polarisation of the scattered quark or the
target remnant
is transferred to the produced hyperon. Therefore, the value of hyperon
polarisation will depend on the spin dynamics in the hadronisation
not only of the scattered quarks, but also of the target remnant. For
the kinematic conditions of the COMPASS experiment the role of these effects has been considered
in \cite{Ell.02}. It occurs that even in the current
fragmentation region the spin transfer from the polarised target
remnant to the \la~hyperon is substantial. The situation is
different for the spin transfer to \al.~ A recent analysis
\cite{Ell.07} shows that at the energy of the COMPASS experiment
the \al~
polarisation is dominated by the spin transfer from the $\bar{s}$-quarks, and
hence is sensitive to
 the $\bar{s}(x)$ distribution in the nucleon.


The polarisation of the produced hyperons depends also on their
spin structure. For example, in the SU(6) quark model the whole spin
of the \la~is carried by the $s$-quark. Therefore, under the assumption of the quark helicity conservation,  even if the struck
$u$- or $d$-quark were completely polarised this would not lead to a polarisation of
the \la~hyperon.
As a consequence, if
the dominant mechanism
of \la~production is the independent $u$- and $d$-quark
fragmentation, then the polarisation of the directly produced \la~hyperons should be
$P_{\Lambda} \sim 0$.

 Using SU(3)$_\mathrm{f}$ symmetry and
experimental data for spin-dependent quark distributions in
the proton, the authors of \cite{BJ} predict that the contributions of $u$- and $d$-quarks
to the \la~spin are negative and substantial, at the level
of 20\% for each light quark. In this model the  spin transfer from $u$-
or $d$-quarks  would lead to a negative spin transfer to \la.

 The \la~and \al~hyperons  can also be
produced  indirectly, via decays of heavy hyperons such as $\Sigma^0, \Sigma(1385), \Xi$ etc.
In this case a non-zero spin transfer from $u$- and $d$-quarks to \la~is also
possible \cite{BGH},\cite{GH.93}.


The spin transfer to \la~and \al~hyperons
has been studied extensively in a number of theoretical models~
\cite{Ans.00,Ash.99,deF.98,Kot.98,Liang-la,Liang-al,Liang09,Ma.00,Yan.00}
for $e^+ e^-$ collisions and lepton DIS.
From a theoretical point of view the simplest case is
 the  polarisation of the hyperons formed in
$e^+e^-$ annihilation at the $Z$-peak,  where hadronisation of the
$q \bar q$-system can be well described using independent
fragmentation. The electroweak theory predicts
for the $s$-quarks from $Z$ decay a longitudinal polarisation of -0.94. The corresponding antiquarks
have the same degree of polarisation, but with opposite helicity.
The fraction of this polarisation that is transferred
to the produced
\la(\al)~hyperon was studied
using $e^+e^-$ annihilation by
ALEPH \cite{ALEPH.96} and OPAL \cite{OPAL.98} experiments at LEP.
Both experiments find large and negative values for
the hyperon longitudinal polarisation, with a strong dependence
on the fraction of the primary quark momentum
carried by the hyperon.
These results
clearly indicate that the fragmentation processes preserve a
strong correlation between the fragmenting quark helicity and the
final hyperon polarisation. However, as was shown in \cite{deF.98}, these data are insufficient to distinguish between predictions of the SU(6) or the BJ \cite{BJ} models.

The longitudinal spin transfer to \la(\al)~in  DIS was
measured in a number of experiments
\cite{WA21,WA59,E632,NOMAD,NOMAD-al,E665,HERMES-lambda,STAR}.  The
earlier neutrino DIS experiments \cite{WA21,WA59,E632} have found
an indication for a large negative polarisation of \la~hyperons
formed in the target fragmentation region.
However, the statistics of each of these experiments does not exceed 500 events.
The E665 Collaboration \cite{E665} has measured
\la(\al)~production using 470~GeV positive muons scattered off
hydrogen, deuterium and other nuclear targets. The total
number of events amounts to  750 \la~and 650 \al.
The spin transfers of \la~and \al~hyperons were found to have opposite signs: negative for \la~and positive for \al.
The NOMAD Collaboration \cite{NOMAD,NOMAD-al} has studied  \la~
and \al~ polarisation in DIS using 43~GeV muon neutrinos. The
total number of events amounts to 8087 \la, mainly in the target fragmentation region, and 649 \al.
The
results
confirm those
of  earlier experiments \cite{WA21,WA59,E632}
 and show
 a large negative longitudinal polarisation $P_{\Lambda}$
in the target fragmentation region
while no
%
significant spin transfer to \la~was detected in the
current fragmentation region.
This region
was later explored by the HERMES Collaboration \cite{HERMES-lambda}.
The \la~polarisation was measured using the 27.6~GeV longitudinally polarised
positron beam. The total statistics is 7300 \la.
Within the experimental uncertainties, the resulting spin transfer was found to be compatible with zero.
The STAR Collaboration
at RHIC \cite{STAR} has measured the longitudinal spin transfer to
\la~and \al~in  polarised proton-proton collisions at a centre of
mass energy $ \sqrt{s} = 200$~GeV. The data sample comprises
 30000 \la~ and 24000 \al. The measurement is limited to
the mid-rapidity region ($\lvert
\eta \rvert<1$) with an average $x_F= 7.5\cdot10^{-3}$. (The Feynman variable is
$x_F= 2p_{L}/W$, where $p_{L}$ is the particle longitudinal momentum in the
hadronic centre-of-mass system, whose invariant mass is $W$.)
 A small
spin transfer, compatible with zero, was found for both  \la~and \al.

The statistical accuracy achieved in the measurements discussed above
is quite limited, particularly, in the current fragmentation region.
The data on the polarisation of the \la~ hyperons  tend to be compatible with
zero and no clear conclusions about the spin transfer mechanism can
be drawn. The dependence of the spin transfer  on the Bjorken scaling variable $x$ and
$x_F$ has been mapped out by the HERMES experiment \cite{HERMES-lambda},
but for the \la~hyperon only.

The present analysis is based on about 70000 \la~and 42000 \al~events.
Our data
allow to explore the $x$-dependence of the spin transfer to \la~in
a large $x$-interval and to measure, for the first time, the $x$-
and $x_F$-dependences of the spin transfer to \al.
A substantial \al~polarisation is found, which is important for the
investigation of the strange quark distribution in the nucleon.


\section{\label{sec:cuts} Data analysis}

We have studied \la~ and \al~  production by scattering 160~GeV
polarised $\mu^+$ off a polarised $^{6}$LiD target in 
 the COMPASS experiment (NA58) at CERN.
 A detailed description of the COMPASS experimental setup
is given elsewhere \cite{COMPASS}.

The data used in the present analysis were collected during the
years 2003--2004. The longitudinally polarised muon beam
has  an average polarisation of $P_b=-0.76\pm0.04$ in the 2003 run and of
$P_b=-0.80\pm0.04$ in the 2004 run. The momentum of each beam muon is measured upstream of the experimental area in a beam momentum station consisting of several planes of scintillator strips or scintillating
fibres with a dipole magnet in between.
The target consists of two 60 cm long,
oppositely polarised cells.
The data from both longitudinal target spin orientations were
recorded simultaneously and averaged in the present analysis.
The number of events with \la(\al)~hyperons for
each target spin orientation is the same within 1\% accuracy.



The event selection requires a reconstructed
interaction vertex defined by the incoming and the scattered muon
located inside the target.
DIS events are selected by cuts on the photon virtuality ($Q^{2}
>1~$~(GeV/{\it c})$^{2}$) and on the fractional energy of the virtual photon ($0.2<y<0.9$).
The data sample 
consists of
$8.67\cdot10^7$ DIS events from the 2003
run and $22.5\cdot10^7$ DIS events from the 2004 run.

The   \la~and  \al~hyperons  are identified
by their  decays into $p\pi^-$ and $\bar{p}\pi^+$. To estimate
systematic effects, decays of $K^{0}_{S} \to \pi^+\pi^-$ are
also analysed.
 Events with \la, \al~and \ks~
 decays are selected by demanding
that two hadron tracks form a secondary vertex.
Particle identification
provided by a ring imaging Cherenkov detector and  calorimeters
is not used for hadrons in the present analysis.
The results of the analysis for RICH-identified hadrons will be the subject of a separate paper with larger statistics.
In order  to suppress background events,
the secondary vertex is required to  be within a 105 cm long
fiducial region starting 5 cm downstream of the
target.
The angle $\theta_{col}$ between the  hyperon momentum and the
line connecting the  primary and the secondary vertex is required to be
$\theta_{col}<0.01$ rad.
This cut selects events with the correct direction of the hyperon
momentum vector with respect to the primary vertex,
which results in
a reduction of the combinatorial background.
 A cut on the transverse momentum $p_t$
of the decay products with respect to the hyperon direction of $p_t
>23$ MeV/c is  applied to reject $e^{+}e^{-}$ pairs due to
$\gamma$ conversion.
Only particles with momenta larger than 1~GeV/c were selected
to provide optimal tracking efficiency.



\begin{figure*}[htb] 
\begin{center}
\begin{tabular}{cc}
\epsfig{file=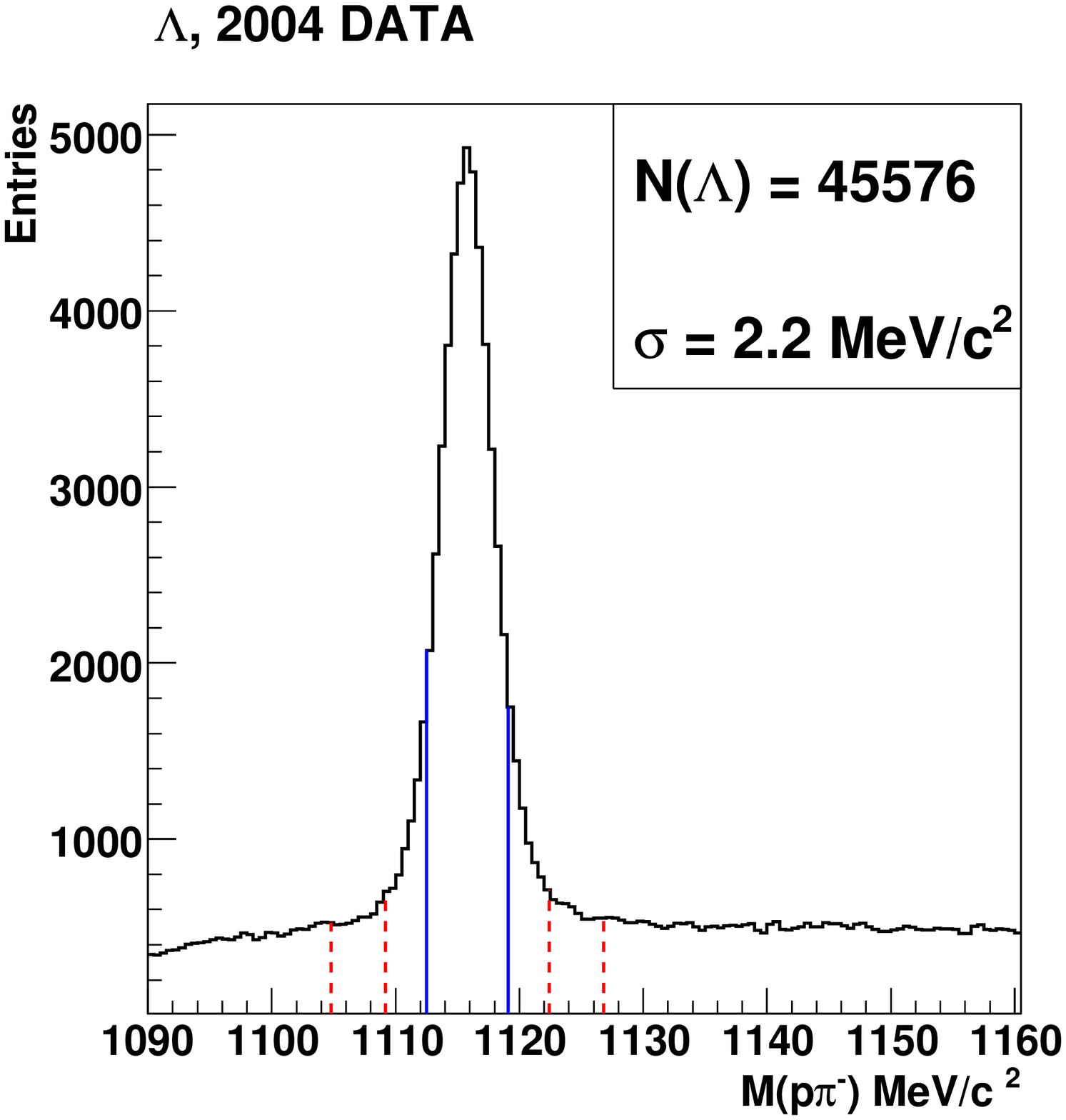,width=6cm} &
\epsfig{file=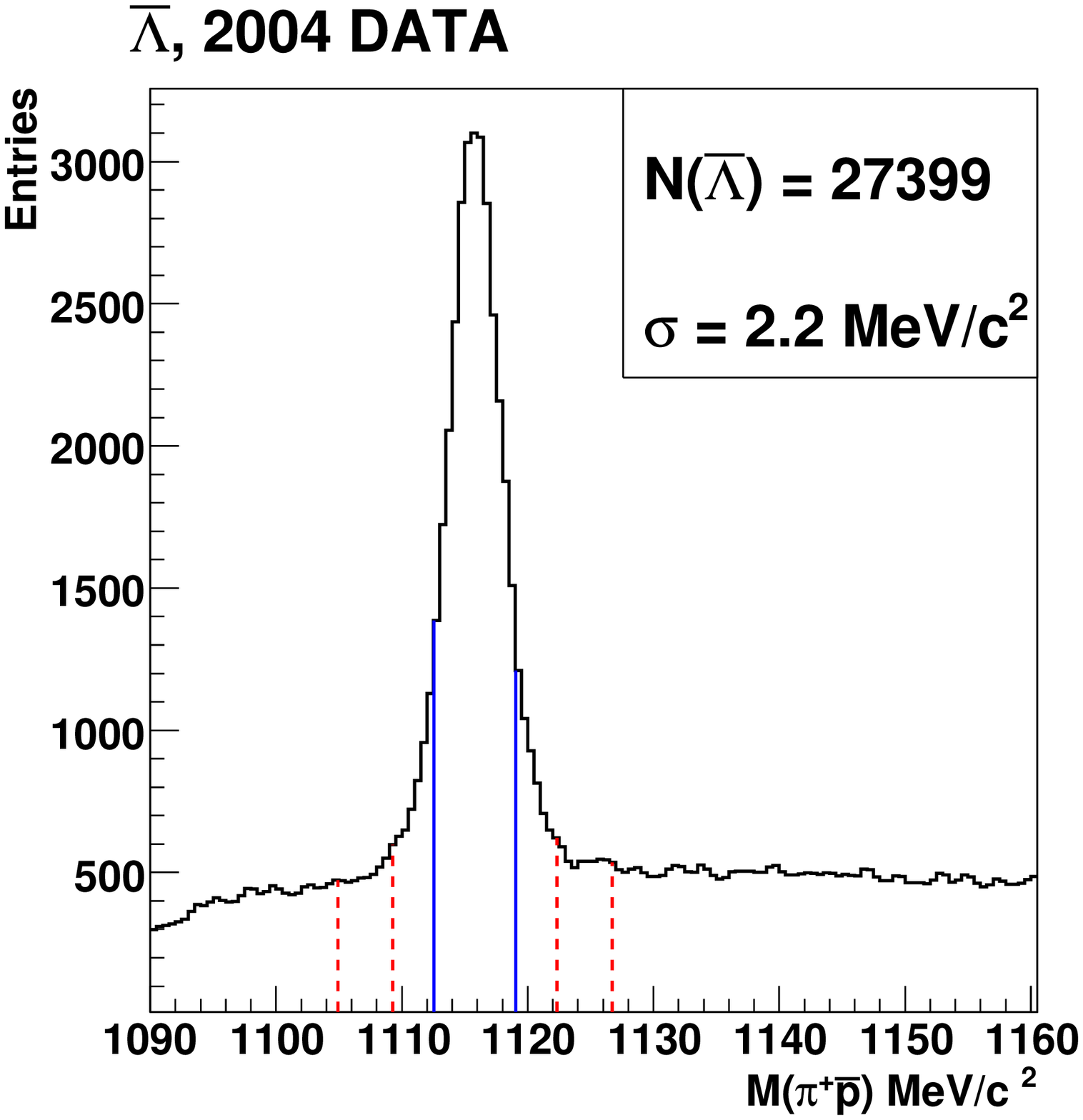,width=6cm} \\
a) & b) \\
\end{tabular}
\caption{The invariant mass distribution for the $p\pi^-$ (a) and
$\bar{p}\pi^+$ (b) hypothesis for the data of 2004 run. The solid
lines marks the band of the \la(\al)~signal, the dashed lines
show the sidebands used for determination of the
background regions.} \label{inv-mass}
\end{center}
\end{figure*}

The  $p\pi^-$ and $\bar{p}\pi^+$
invariant mass distributions with peaks of \la~and \al~
are shown in Fig.~\ref{inv-mass} for
the data of the 2004 run.
The
main sources of background are events from  \ks~ decays
and combinatorial background. A Monte Carlo (MC) simulation
shows that the percentage of  kaon
background changes from 1 to 20 \%, when  $\cos \theta$ varies
from -1 to 1, where $\theta$ is the angle between the direction of the
decay proton(antiproton) in the \la(\al)~rest frame and the quantisation axis along the
momentum vector of the virtual photon.
At small and negative values of $\cos \theta$ the invariant mass distribution of the kaon background is flat. However, at $\cos \theta \sim 1$,
the kaon contribution is
concentrated mainly at small values of the invariant mass on the left side of the \la(\al)~peak in Fig.~\ref{inv-mass}. In this angular region the distribution of the kaon events under the \la(\al)~peak changes rapidly. In order to minimise the influence of the
kaon background, the angular interval was limited to
$-1<\cos{\theta} < 0.6$ (a similar cut was introduced in the analysis of the STAR data \cite{STAR}). This cut reduces the \la(\al)~signal by $\sim$10\%.


 The total number of events after all
selection cuts is $N(\Lambda)=69500 \pm 360$, $N(\bar{\Lambda)}=41600 \pm 310$ and
$N(K^0_S)=496000\pm 830$.
The large amount of \al~events
is a unique feature of the COMPASS experiment.


\section{\label{sec:results} Experimental results}

\begin{figure*}[htb] 
\begin{center}
\begin{tabular}{cc}
\epsfig{file=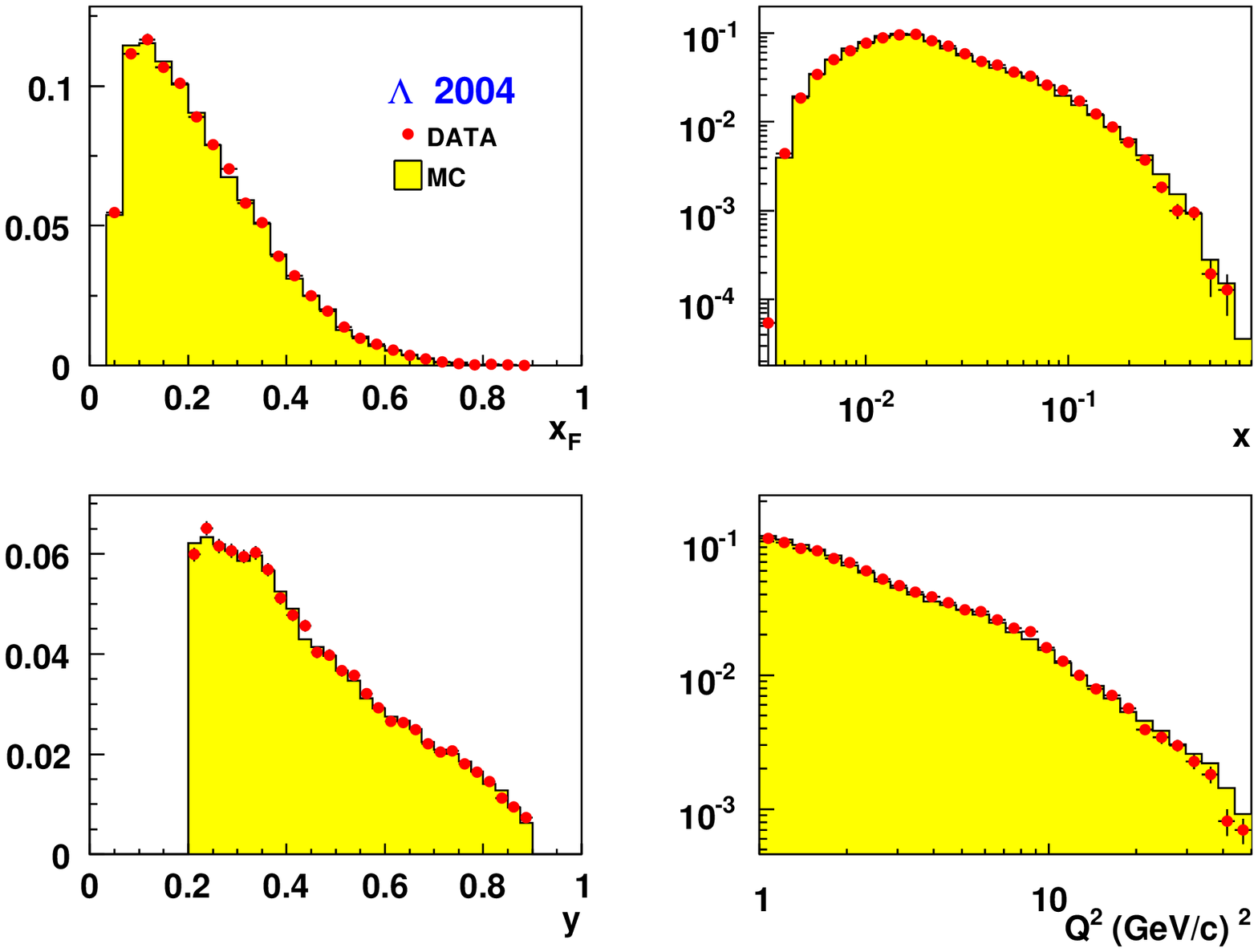,width=7cm} &
\epsfig{file=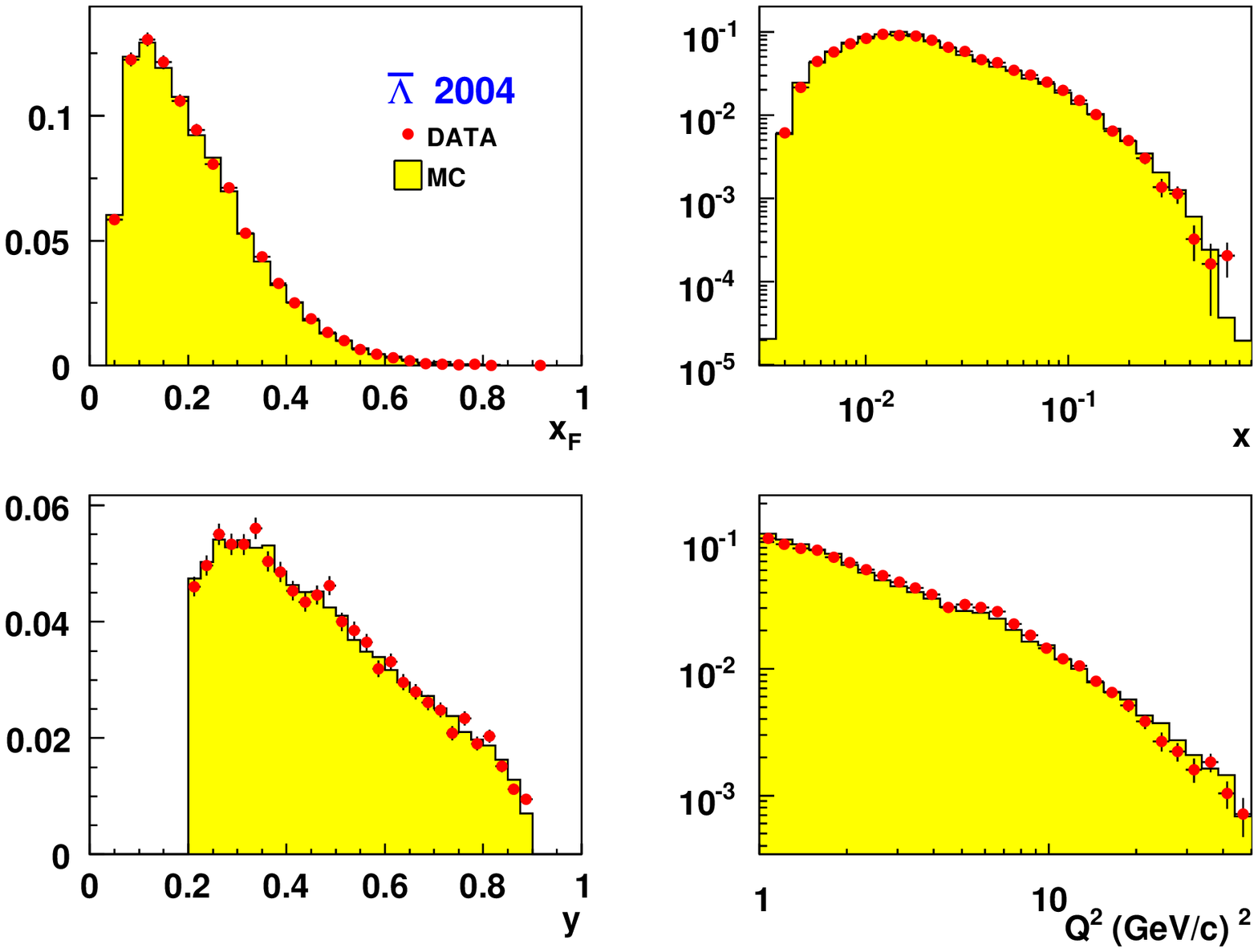,width=7cm} \\
a) & b) \\
\end{tabular}
\caption{Comparison between the experimental data (circles) and MC
(histograms) on $x_F, x, y$ and $Q^2$ distributions of the \la~
(a)
 and \al~ (b) hyperons. The data are for the 2004 run.} \label{kin}
\end{center}
\end{figure*}

The distributions of experimental and MC events satisfying the
selection cuts are shown in Fig.~\ref{kin} for \la~and \al~as functions of different kinematic variables for events of the 2004 run. The data from the 2003 run have similar distributions. Both experimental
and MC distributions are normalised on the total number of events.
The  acceptance of the COMPASS spectrometer selects
\la(\al)~only in the current fragmentation region. For this analysis we
take \la(\al)~events in the interval  $0.05<x_F<0.5$ with the
average value $\bar{x}_F=0.22~(0.20)$. In contrast to other DIS
experiments \cite{NOMAD,NOMAD-al,E665,HERMES-lambda},
our data cover a large region of Bjorken $x$ ($\bar{x}=0.03$)
extending to values of  $x$ as low as $x=0.005$.
The average value of \la~ fractional energy
is $\bar{z}=0.27$.
 The average values of $y$ and $Q^2$ are 0.46 and 3.7 ${\rm (GeV/c)^2}$, respectively.

The distributions of the kinematic variables for  \al~produced
in  DIS  differ from  the \la~ones, since \al~production is
suppressed in the target fragmentation region. However, in the
current fragmentation region
the \al~have
practically the same kinematic distributions as the \la~(see
Fig.~\ref{kin}).

The shaded histograms in Fig.~\ref{kin} show the same
distributions for  MC events. The COMPASS Monte Carlo
code is based on the LEPTO 6.5.1 generator \cite{LEPTO} providing
DIS events which are passed through a GEANT-based apparatus
simulation programme and the same chain of reconstruction
procedures as the experimental events. To provide better agreement
between data and MC, a tuning of several LEPTO (JETSET) parameters
has been performed.
The  values of the modified parameters are compared with the
default ones \cite{JETSET} and with those used in previous experiments in Table~\ref{FF}.
Here, the PARJ(21) parameter corresponds to the width of the Gaussian transverse momentum
    distribution for the primary hadrons.  The parameters PARJ(23), PARJ(24) are used
    to add non-Gaussian tails to the transverse momentum distribution,
    PARJ(41), PARJ(42) are the parameters of the symmetric Lund
    fragmentation function \cite{JETSET}.

    \begin{table}[!htb]
      \caption{Comparison of the default \cite{JETSET} and modified JETSET parameters.
      Corresponding values used in the HERMES \cite{HERMES-lambda} and NOMAD \cite{NOMAD} experiments are also given.}
 \begin{center}
      \begin{tabular}{lllll}
      \hline\\
      Parameters & Default  & Used  & HERMES & NOMAD  \\
      \hline\\
      PARJ(21)   & 0.36 & 0.4  & 0.38   & 0.41    \\
      PARJ(23)   & 0.01 & 0.08 & 0.03   & 0.15    \\
      PARJ(24)   & 2.0  & 2.5  & 2.5    & 2.0     \\
      PARJ(41) & 0.3  & 0.95 & 1.13   & 1.5     \\
      PARJ(42)  & 0.58 & 0.37 & 0.37   & 0.9     \\
      \hline
      \end{tabular}
      \end{center}
      \label{FF}
    \end{table}

    %
%

Some small but systematic differences were observed between the
momentum spectra seen in the data and those given by the MC simulation.
To overcome these differences, the
reconstructed MC events are weighted in order to provide the same
momentum and $z$  
distributions as  the data.
The weighting of the MC events leads to a slight
modification of the angular acceptance, which
is below 2~\% in
the whole angular range.


\subsection{Determination of the angular distributions}

The acceptance corrected angular distribution of the decay protons(antiprotons) in the
\la(\al)~rest frame is



\begin{equation}
\frac{1}{N_{tot}}\cdot
\frac{dN}{d\cos{\theta}}=\frac{1}{2}\cdot(1+\alpha P_L
\cos{\theta}). \label{real}
\end{equation}
Here, $N_{tot}$ is the total number of
acceptance corrected
 \la(\al), the
longitudinal polarisation $P_L$ is the projection of the polarisation vector
on
the momentum vector of the
virtual photon,
$\alpha=+(-)0.642\pm0.013$ is the \la(\al)~decay parameter,
$\theta$ is the angle between the direction of the
decay proton for \la~(antiproton - for \al, positive $\pi$ - for
\ks) and the corresponding axis. The acceptance correction
was determined using the MC
simulation for unpolarised \la~and \al~decays. The angular dependence of the
acceptance is quite smooth, it decreases by a factor 1.2-1.3 in the angular
interval used.

 To
determine the  \la (\al)~angular distributions, 
a sideband
subtraction method is used. The events with an invariant mass within
a $\pm 1.5~\sigma$ interval from the mean value of the \la (\al)~peak
are taken as  signal.  The background regions are selected
from the left and right sides of the invariant mass peak. Each band
is $2~\sigma$ wide and starts at a distance of $3~\sigma$ from the
central value of the peak. The bands of the signal, as well as the
background regions, are shown in Fig.~\ref{inv-mass}. The \la
(\al)~angular distribution is determined by subtracting the
averaged angular distribution of the events in the sidebands from
the angular distribution of those in the signal region.

\begin{figure*}[htb!] 
\begin{center}
\epsfig{file=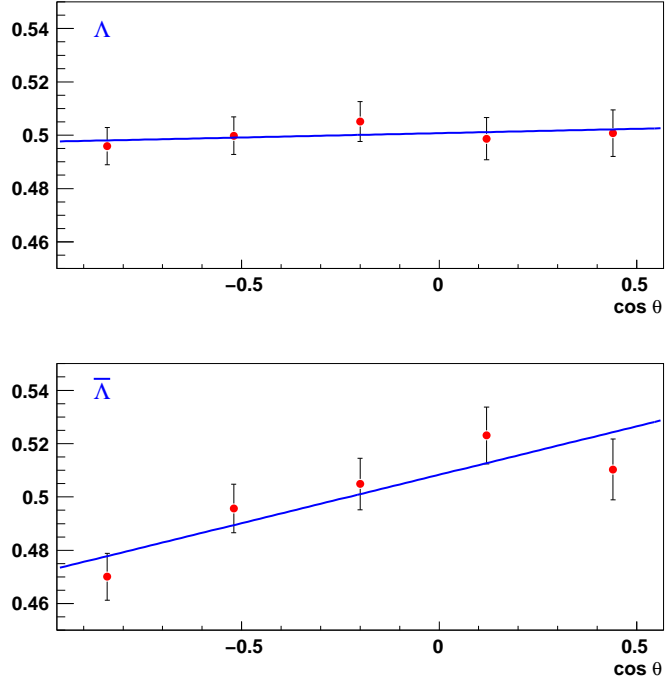,width=10cm,clip=} \caption{The normalised angular
distributions, corrected for acceptance, for the
\la~and \al~for
all events of the 2004 run. The solid line is the
result of the linear fit to Eq.~(\ref{real}).} \label{ang-dis}
\end{center}
\end{figure*}

Figure~\ref{ang-dis} shows the acceptance corrected angular
distributions
 for all events of the 2004 run for
 \la~and
 \al. The events of the 2003 run have similar
 angular distributions.

\subsection{Longitudinal spin transfer}






%

The spin transfer coefficient $D_{LL'}^{\Lambda}$
describes the probability that the
polarisation of the struck quark along the primary quantisation axis $L$ is transferred to the \la~hyperon along the secondary quantisation axis $L'$. In our case the primary and the
secondary axes are the same $L=L'$ and coincide with the virtual
photon momentum.
The longitudinal spin transfer relates the longitudinal
polarisation of the hyperon $P_L$ to the polarisation of the
incoming lepton beam $P_b$:

\begin{equation}
P_L= D_{LL}^{\Lambda}\cdot P_b \cdot D(y).    \label{s}
\end{equation}
The longitudinal polarisation $P_L$ is determined from a fit of
the angular distribution (\ref{real}). The virtual photon depolarisation factor $D(y)$ is given by:

\begin{equation}
 D(y) =\frac{1-(1-y)^2}{1+(1-y)^2}.     \label{D}
\end{equation}
To evaluate the spin transfer, the product $P_b \cdot D(y)$ is
calculated for each event passing the selection criteria.
The beam polarisation $P_b$ is parametrised as a function of the incident muon momentum \cite{COMPASS}.
The $P_b \cdot D(y)$ distribution is determined by subtraction of the averaged
distribution of the sideband events from the distribution of the events
in the signal region, marked by solid lines in Fig.~\ref{inv-mass}.
The
average value of the $P_b \cdot D(y)$ distribution is used to
calculate the $D_{LL}^{\Lambda}$ according to Eq.~(\ref{s}).





The weighted averages of the spin transfers for the 2003 and the 2004 data
are:


\begin{eqnarray}
D_{LL}^{\Lambda}=-0.012\pm0.047(stat)\pm0.024(sys),~\bar{x}_F=0.22 \label{sla1}, \\
D_{LL}^{\bar{\Lambda}}=0.249\pm0.056(stat)\pm0.049(sys),~ \bar{x}_F=0.20. \label{sal1}
\end{eqnarray}

The systematic errors are mainly due to the uncertainty
of the acceptance correction determined by the Monte
Carlo simulation. The presence of possible systematic effects
was checked by looking at the result of a physical process with no polarisation
effects, i.e. the longitudinal spin transfers to the \ks~
and by checking the stability of the result by varying the selection cuts.
The longitudinal
      spin transfer to the \ks~ turns out to be $D_{LL}^{K^0_S}=0.016\pm0.010$.
The value of the longitudinal spin transfer
for kaons is taken as an estimate for the corresponding systematic error $\delta(K^0_S)$.
Some systematic effect
$\delta(\theta)$ appears due to variation of the cut on $\cos \theta$.
The uncertainty of the sidebands subtraction method, $\delta(ss)$ is estimated by varying the width of the central band
      of Fig.~\ref{inv-mass} from $\pm 1.5\,\sigma$ to $\pm 1$, $\pm 1.25$,
      $\pm 1.75$ and $\pm 2\,\sigma$.
      Another source of systematic errors is the uncertainty in the
beam polarisation, $\delta(P_b)$. The relative error in the value
of beam polarisation is 0.05.

    The values of the systematic errors are given in Table \ref{syst}.
The total systematic error was obtained by summing
the various contributions in quadrature.

\begin{table}[htb!]
\caption[6s]{Systematic errors for the spin transfer to \la~ and \al.\\}
\begin{tabular}{lrr}
\hline\\
  & \la  & \al  \\
\hline\\
Spin transfer to kaons, $\delta(K^0_S)$ & 0.016 &  0.016         \\
Variation of the $\cos \theta$ cut, $\delta(\theta)$ & 0.016 &   0.044        \\
Uncertainty of the ss-method, $\delta(ss)$ & 0.010 &   0.016        \\
Uncertainty of the beam polarisation, $\delta(P_b)$ & 0.0006 &   0.013        \\
\hline \\
$\sigma_{syst}$ & 0.024 & 0.049 \\
\hline
\end{tabular}
\label{syst}
\end{table}

\subsection{Dependence of the \la~ and \al~ spin transfer on $x$ and $x_F$ }

\begin{figure*}[htb!] 
\begin{center}
\epsfig{file=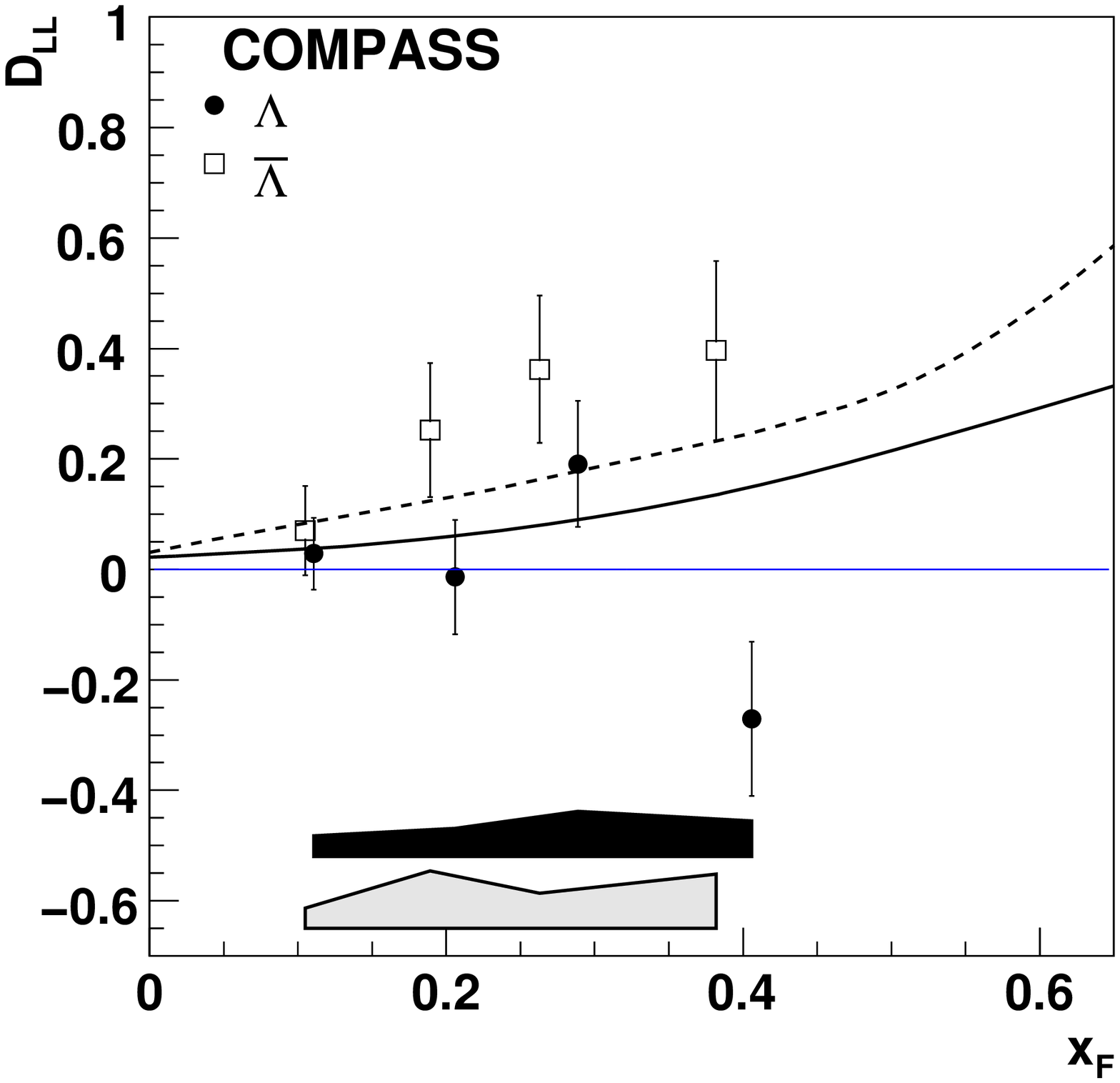,width=10cm,clip=} \caption{The $x_F$
dependence of the longitudinal spin transfer to $\Lambda$ and
$\bar\Lambda$. The solid line corresponds
to the theoretical calculations of
\cite{Ell.07}(Model B, SU(6),CTEQ5L) for \la~ and the dashed line is for the \al~spin
transfer. The shaded bands show the size of the
    corresponding systematic errors.} \label{xf}
\end{center}
\end{figure*}

The $x$ and $x_F$ dependences of the spin transfers to \la~and
\al~are shown in Figs.~\ref{xx} and \ref{xf}.
These dependences are different for \la~and \al.
The spin
transfer to \la~is small and compatible
with zero in the entire $x$ range, while
the spin transfer to \al~may reach values as large as $D_{LL}^{\bar{\Lambda}}=0.4-0.5$.

A similar difference between \la~and \al~spin transfers is
observed in the $x_F$ dependence (Fig.~\ref{xf}).
The spin transfer to \al~
tends to increase with $x_F$, while the
\la~ one
does not show any
 significant $x_F$ dependence.


\section{\label{sec:discussions} Discussion of the results}

\begin{figure*}[htb!] 
\begin{tabular}{cc}
\epsfig{file=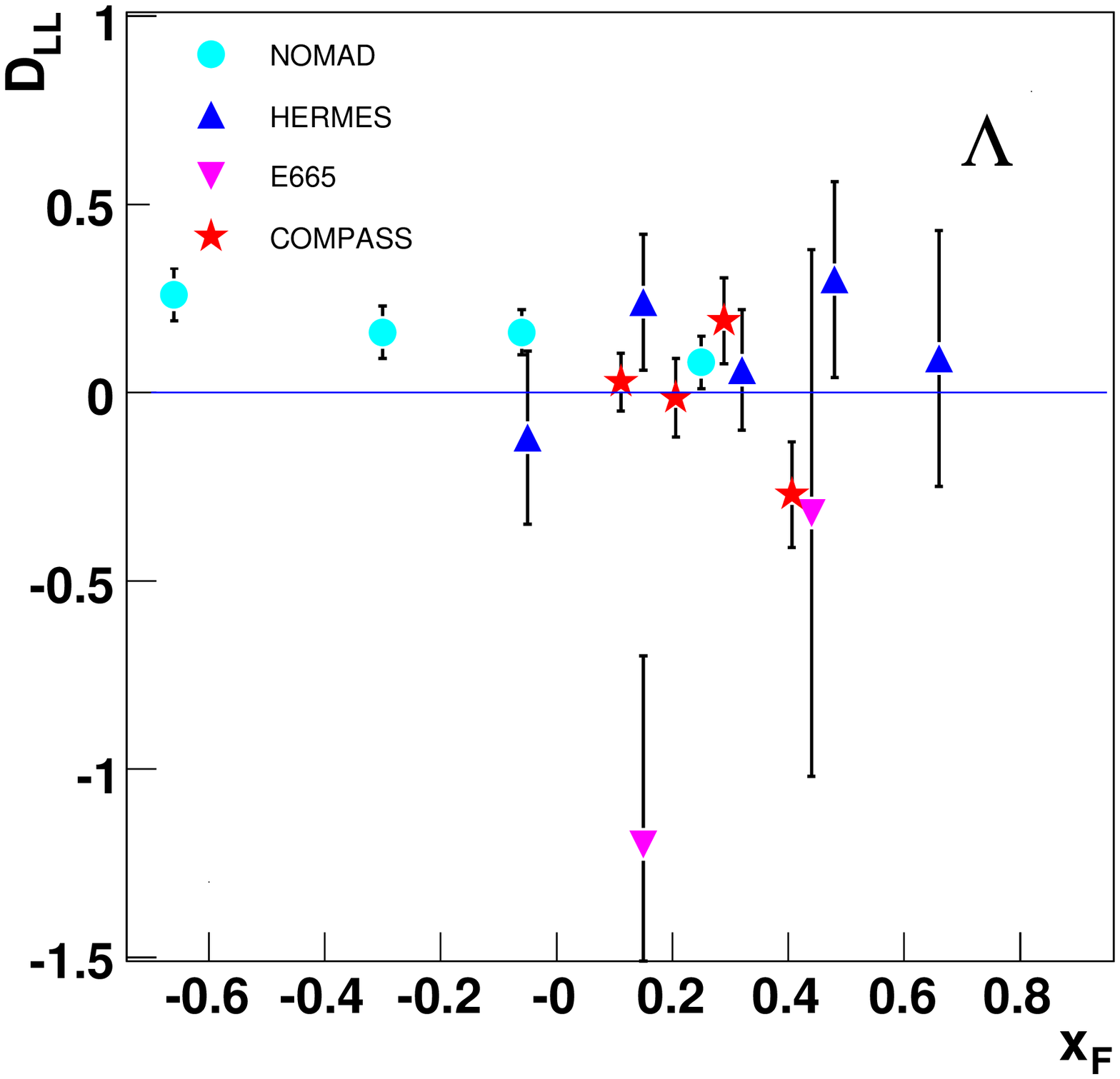,width=7cm,clip=} &
\epsfig{file=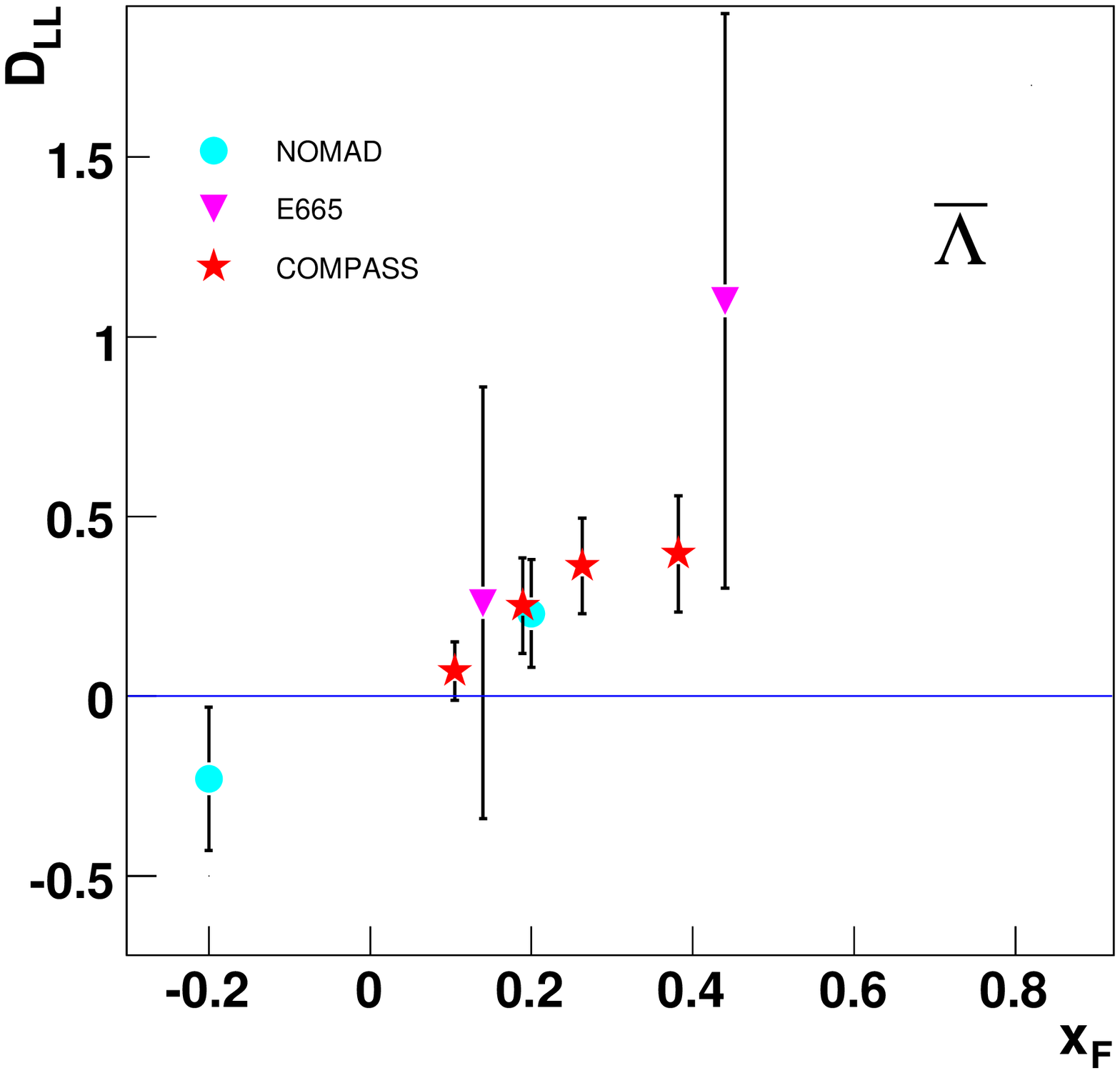,width=7cm,clip=} \\
a) & b) \\
\end{tabular}
\caption{The $x_F$ dependence of the longitudinal spin transfer
to  \la~(a) and \al~(b) for the COMPASS (stars) and other
experiments \cite{NOMAD,NOMAD-al,E665,HERMES-lambda}(NOMAD data-
circles, E665 - reverse triangles, HERMES - triangles).}
\label{comp}
\end{figure*}

A comparison of the  $x_F$ dependence of the longitudinal spin transfer to \la~and
\al~for COMPASS and other experiments \cite{NOMAD,NOMAD-al,E665,HERMES-lambda} is shown  in Fig.~\ref{comp}. 
For \la~there is general agreement between the present results and existing data.
For \al~ the measurement of the E665 Collaboration \cite{E665}  indicated
a positive spin transfer (see
 Fig.~\ref{comp}b).
The present
result
 confirms this observation
with a much better statistical precision.
The NOMAD Collaboration has found \cite{NOMAD-al} that the
spin transfer to \al~ is $D_{LL}^{\bar{\Lambda}}=0.23\pm0.15\pm0.08$
at $\bar{x}_F=0.18$
in a good agreement with the present result (\ref{sal1}).
The measured  $D_{LL}^{\bar{\Lambda}}$
increases with $x_F$, the same trend
was found for the
\la~polarisation in the experiments at LEP \cite{ALEPH.96,OPAL.98}.

The main conclusion from the our results
is that the longitudinal spin transfers to \la~and \al~hyperons in
DIS are not equal. To understand this phenomenon let us consider
the leading order (LO) parton model, where
the spin transfer to
\la(\al)~produced on an unpolarised target
by  polarised leptons
is given by (see for example \cite{Kot.98}):



\begin{equation}
 D_{LL}^{\Lambda(\bar{\Lambda})}(x,z)=\frac{\sum_{q} e^2_q  q(x)
\Delta D^{\Lambda(\bar{\Lambda})}_q(z)}{\sum_{q} e^2_q  q(x) D^{\Lambda(\bar{\Lambda})}_q(z)}.
\label{plam}
\end{equation}
Here $e_q$ is the quark charge,
 $q(x)$ is the
unpolarised  quark distribution function,
$D^{\Lambda(\bar{\Lambda})}_q(z)$ and $\Delta D^{\Lambda(\bar{\Lambda})}_q(z)$ are the unpolarised
and the polarised quark fragmentation functions.


Practically all models of the \la~spin structure predict that the
contribution from the $s$-quark to the \la~spin is dominant. This contribution
varies from 100\% for the SU(6) model to 60-70\% in the BJ-model \cite{BJ} or the
lattice-QCD calculation \cite{Goc.02}.
It means that  scattering off $u$- or $d$-quarks is important for the \la(\al)~production but not
for the spin transfer to \la(\al). Accordingly, the polarised fragmentation functions
$\Delta D_s^{\Lambda}$
entering in the numerator of Eq.~(\ref{plam}) is expected to be much larger than its light quarks
counterparts. Therefore, one may assume that
$\Delta D^{\Lambda}_q(z)=~~ \allowbreak\Delta D^{\bar\Lambda}_{q}(z)\sim0$
for $q=u,d,\bar u, \bar d$.
Then the sum in the
numerator of (\ref{plam}) is reduced to the
strange quark contribution:

\begin{eqnarray}
 D_{LL}^{\Lambda}(x,z) \approx \frac{1}{9}\frac{s(x) \Delta D^{\Lambda}_s(z)}
 {\sum_{q} e^2_q  q(x) D^{\Lambda}_q(z)}, \label{plam1}\\
 D_{LL}^{\bar{\Lambda}}(x,z) \approx \frac{1}{9}\frac{\bar s(x) \Delta D^{\bar{\Lambda}}_{\bar s}(z)}
 {\sum_{q} e^2_q  q(x) D^{\bar{\Lambda}}_q(z)}.\label{plam2}
\end{eqnarray}

From (\ref{plam1})-(\ref{plam2}) it
is seen that the spin transfer from the polarised lepton to \la~and
\al~must be different even if  $s(x)=\bar{s}(x)$.
The reason is that
the denominators
of (\ref{plam1}) and (\ref{plam2})
 are
 proportional to the \la(\al) production cross section. Due to the combined effect of
 the $u$-quark dominance and  the favoured fragmentation of  $u$-quark to \la, as opposed
 to \al, the cross section for \la~is expected to
 be larger.  This expectation is confirmed by the measured yields of \la~and
 \al, reported in Sect. 2.
 Therefore, one may expect
a smaller spin transfer to \la~than to \al.

\begin{figure*}[htb!] 
\begin{center}
\epsfig{file=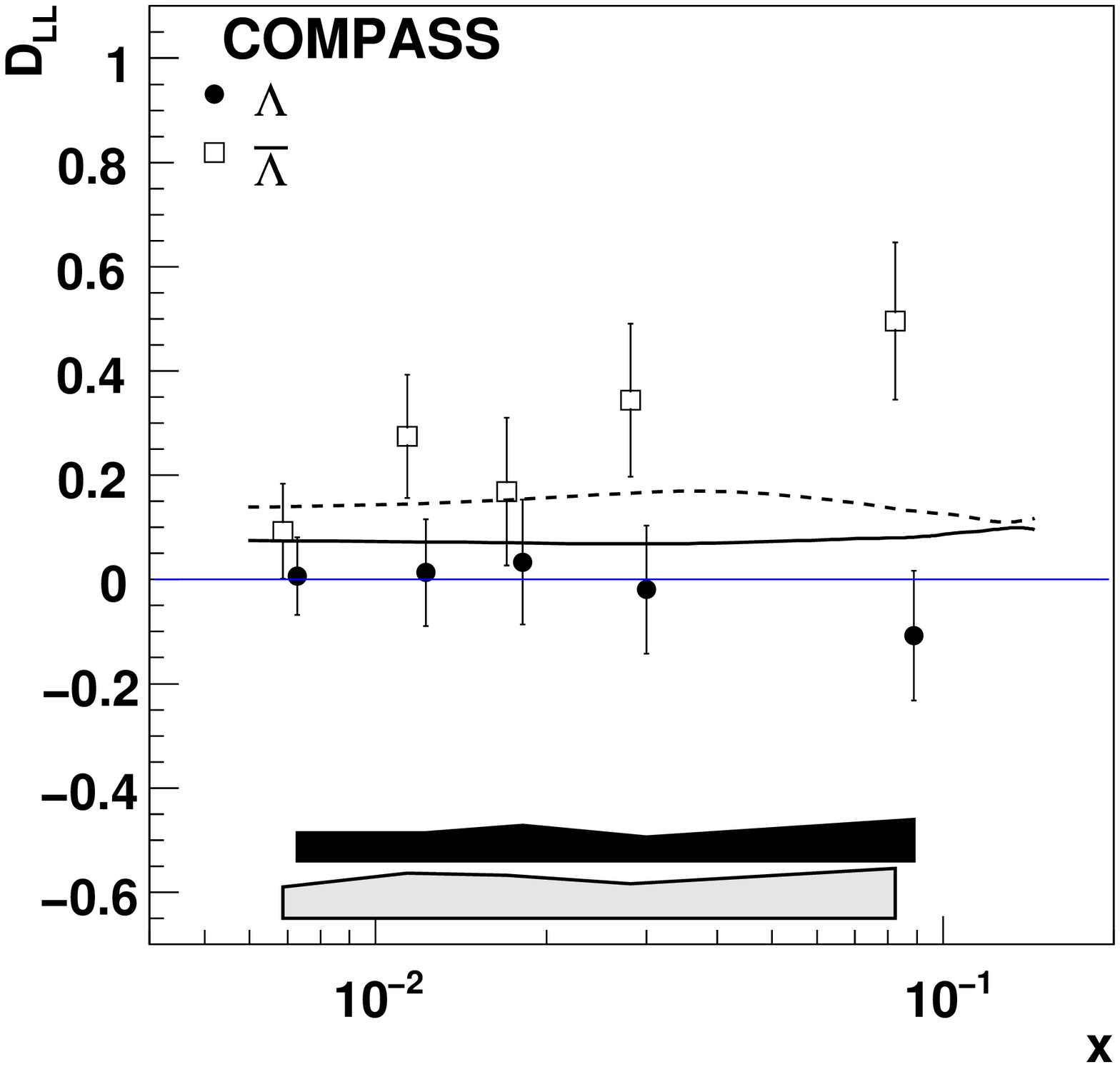,width=10cm,clip=} \caption{The $x$
dependence of the longitudinal spin transfer to \la~ and
\al. The solid line corresponds
to the theoretical calculations of
\cite{Ell.07}(Model B, SU(6),CTEQ5L) for \la~ and the dashed line is for the \al~spin
transfer.The shaded bands show the size of the
    corresponding systematic errors.} \label{xx}
\end{center}
\end{figure*}

This conclusion of the LO parton model has been confirmed by
the calculation of  \cite{Ell.07},
shown in Figs.~\ref{xx},\ref{xf}. The model is based on the LEPTO \cite{LEPTO} MC
event generator, in which  the independent fragmentation is replaced by the
hadronisation of the string formed by the struck quark
 and the
target remnant.  All contributions, including
those from the target remnant and from decays of heavy hyperons
are taken into account.
Figs.~\ref{xx},\ref{xf}
    show that indeed calculations of
the model
\cite{Ell.07} lead to a larger
spin transfer to \al~than
to \la. The same trend was found in
the recent calculation of  \cite{Liang09}.

\begin{figure*}[htb] 
\begin{tabular}{cc}
\epsfig{file=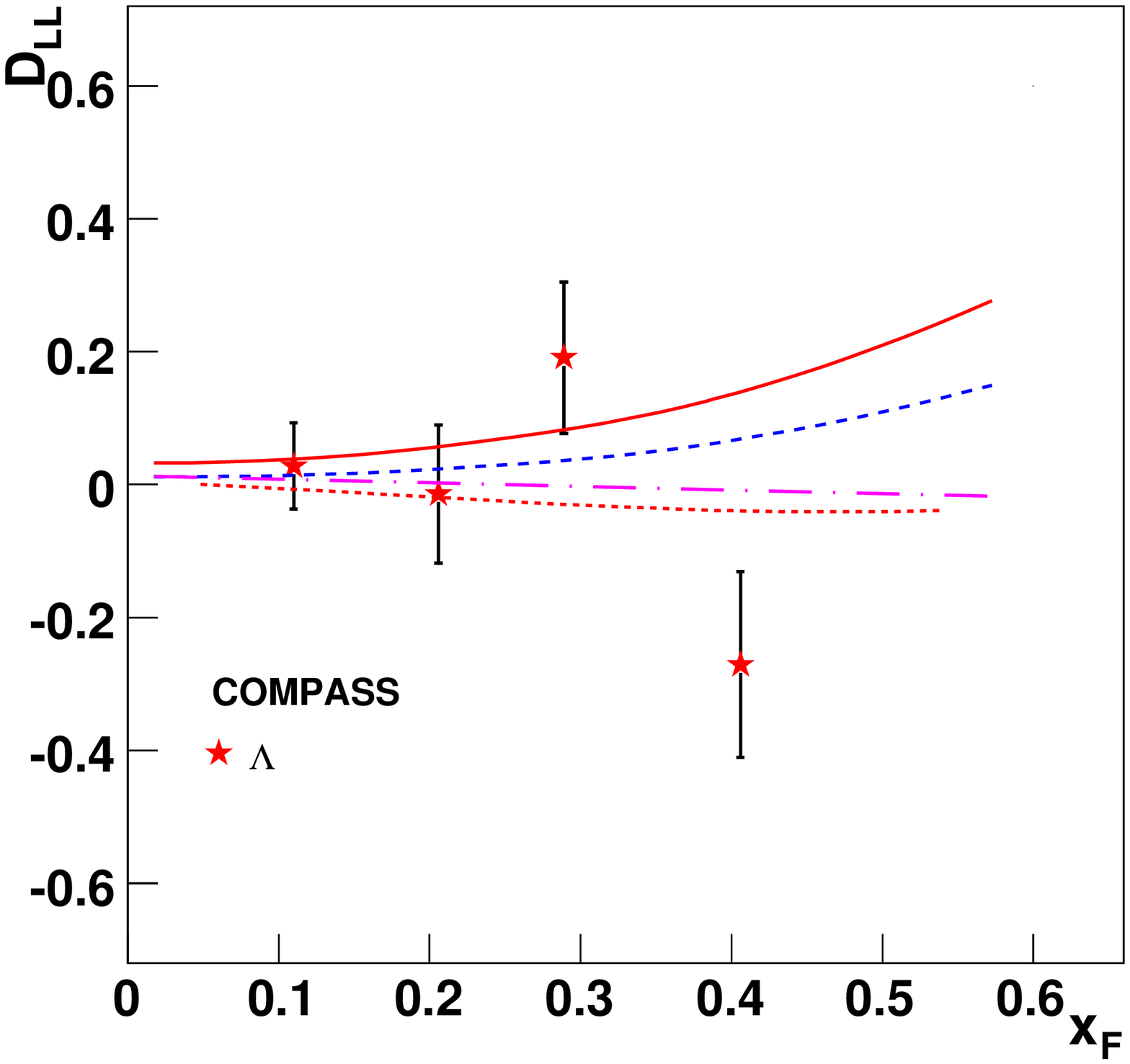,width=7cm} &
\epsfig{file=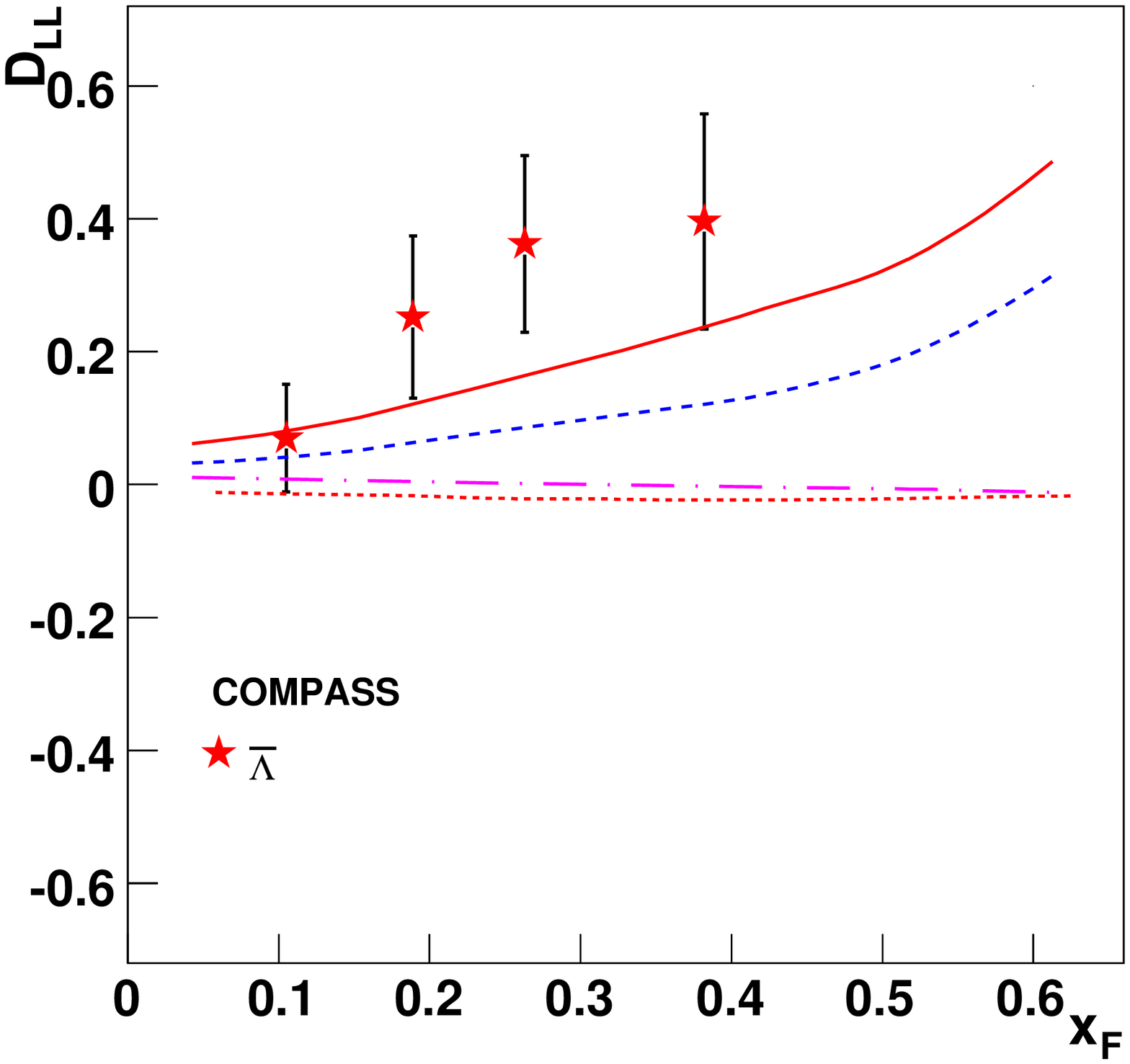,width=7cm} \\
a) & b) \\
\end{tabular}
\caption{The $x_F$ dependences of the longitudinal
spin transfer to  \la~(a) and \al~(b) calculated in \cite{Ell.07}(model B)
for the GRV98LO parton distribution functions (dashed
lines), the CTEQ5L pdf (solid lines) and for the CTEQ5L without spin transfer from the
$s$-quark (dash-dotted lines). The SU(6) model for the \la~spin structure is assumed.
The dotted lines corresponds to the calculations for the CTEQ5L without spin transfer from the
$s$-quark in the BJ-model of \la~spin \cite{BJ}.}  \label{comp-pdf}
\end{figure*}

Another indication from the parton model Eqs.~(\ref{plam1})-(\ref{plam2})
is that the contribution from
the strange quarks (antiquarks) is essential for the spin transfer to \la(\al).
This observation is also confirmed by the results of \cite{Ell.07}.
In Fig.~\ref{comp-pdf} the degree of the sensitivity to the
strange parton distributions
is illustrated by the comparison of the results obtained with
the CTEQ5L \cite{CTEQ}(solid line) and GRV98LO \cite{GRV98}(dashed line)
parton distributions.
The GRV98 set is chosen because of its assumption that there is no
intrinsic nucleon strangeness at a low scale and the strange sea
is of pure perturbative origin. The CTEQ Collaboration allows
non-perturbative  strangeness in the nucleon. The amount of this
intrinsic strangeness is fixed from the dimuon data of the CCFR
and NuTeV experiments~\cite{CCFR}. As a result, the $s(x)$
distribution of CTEQ is larger than the GRV98 one by a factor of
about two in the region $x=0.001-0.01$. The results in
Fig.~\ref{comp-pdf} show that
the data on \la~can not discriminate between the
predictions since the spin transfer to \la~is small.
For the \al~hyperon the  use of CTEQ5L  set leads to a prediction
which is nearly twice larger than the one with the GRV98LO  and much closer to
the data. This behaviour reflects the difference in the corresponding
${\bar s}$-quark distributions.
If one completely switches off the spin transfer from the $s(\bar s)$ quarks,
the spin transfer to \la(\al)~ practically vanishes (dash-dotted line).
This feature is independent of the model of \la~spin structure.  Calculations in the
BJ-model \cite{BJ}, where the spin transfer from the $u$- and $d$-quarks(antiquarks) is possible,
demonstrate the same absence of the spin transfer
 to hyperon without contribution from the $s(\bar s)$-quarks (dotted line).


At present the strange and antistrange quark distributions $s(x)$ and $\bar{s}(x)$  are
directly accessible only
through the measurement and study of dimuon events in
neutrino and antineutrino DIS \cite{NUTEV}.
The spin transfer to
\al~ could provide an additional
experimental information for determination
of strange quark distributions
in the nucleon. To match this goal the present experimental precision must be
increased and the theoretical uncertainties should be clarified. For instance, the parameters of the
model \cite{Ell.07} have been fixed by fitting the NOMAD data \cite{NOMAD} at
comparatively large $x$ and its predictions can be considered as an illustration of the
possible effects.


\section{\label{sec:concl} Conclusions}

The longitudinal polarisation transfer from  polarised muons to
semi-inclusively produced \la~and \al~hyperons has been studied
in deep-inelastic scattering at the COMPASS experiment. The present data
are the most precise measurements to date of the
longitudinal spin transfer to \la~and \al~in DIS.
The results show
that the spin
transfer to \la~is small with $D_{LL}^{\Lambda}=-0.012 \pm 0.047 \pm 0.024$
at $\bar{x}_F=0.22$. The spin transfer to \al~is larger with
$D_{LL}^{\bar{\Lambda}}=0.249\pm 0.056 \pm 0.049$ at $\bar{x}_F=0.20$. These
values are in agreement  with
results of previous measurements
\cite{NOMAD,NOMAD-al,E665,HERMES-lambda}.

We have also measured the $x$ and $x_F$ dependences of the
longitudinal spin transfer which are different for \la~ and \al~
hyperons.
 The spin transfer to  \la~  is small, compatible with
zero, in the entire domain of the measured kinematic variables.
In contrast, the longitudinal spin transfer to
\al~ increases with $x_F$ reaching values of
$D_{LL}^{\Lambda}=0.4-0.5$.
Comparison with theory shows that the spin transfer to the \al~hyperon strongly depends on the antistrange quark distribution $\bar{s}(x)$ and therefore
the precise measurements of the \al~spin transfer
will provide
 useful information about the antistrange quark distribution $\bar{s}(x)$. 

\section*{Acknowledgement}

We gratefully acknowledge the support of the CERN management and
staff, the skill and effort of the technicians of our
collaborating institutes. Special thanks are due to V.~Anosov and
V.~Pesaro for their technical support during installation and
running of this experiment. It is a pleasure to thank J.~Ellis,
Liang Zuo-tang, D.~Naumov and S.~Belostotsky for stimulating
discussions. 

\bibliographystyle{aipprocl}
\bibliography{reff}
\section*{Figures}
\graphicspath{{figures/}}

\end{document}